\documentclass[nojss]{jss}\usepackage[]{graphicx}\usepackage[]{color}
\makeatletter
\def\maxwidth{ %
  \ifdim\Gin@nat@width>\linewidth
    \linewidth
  \else
    \Gin@nat@width
  \fi
}
\makeatother

\usepackage{lmodern}
\usepackage{amsmath,bm,url}

\title{Efficient Bayesian Structural Equation Modeling in \proglang{Stan}}
\Plaintitle{Efficient Bayesian Structural Equation Modeling in Stan}

\author{Edgar C.\ Merkle\\University of Missouri \And Ellen Fitzsimmons\\University of Missouri \And James Uanhoro\\Ohio State University
  \And Ben Goodrich\\Columbia University}
\Plainauthor{Edgar C. Merkle, Ellen Fitzsimmons, James Uanhoro, Ben Goodrich}

\Shorttitle{Efficient SEM in Stan}

\Abstract{Structural equation models comprise a large class of popular statistical models, including factor analysis models, certain mixed models, and extensions thereof. Model estimation is complicated by the fact that we typically have multiple interdependent response variables and multiple latent variables (which may also be called random effects or hidden variables), often leading to slow and inefficient MCMC samples. In this paper, we describe and illustrate a general, efficient approach to Bayesian SEM estimation in \proglang{Stan}, contrasting it with previous implementations in \proglang{R} package \pkg{blavaan} \citep{merros18}. After describing the approaches in detail, we conduct a practical comparison under multiple scenarios. The comparisons show that the new approach is clearly better. We also discuss ways that the approach may be extended to other models that are of interest to psychometricians.}

\Plainkeywords{Bayesian SEM, structural equation models, Stan, JAGS, MCMC, blavaan} 

\Address{
  Edgar C.\ Merkle\\
  Department of Psychological Sciences\\
  University of Missouri\\
  28A McAlester Hall\\
  Columbia, MO, USA 65211\\
  E-mail: \email{merklee@missouri.edu}\\
  URL: \url{http://faculty.missouri.edu/~merklee/}
}

\IfFileExists{upquote.sty}{\usepackage{upquote}}{}
\begin{document}

\section{Introduction}
Structural equation models (SEMs) are commonly used in the social sciences, where it is customary to (attempt to) measure unobservable traits such as cognitive abilities, attitudes, and proficiencies.  Such models
provide a formal way of connecting these unobservable traits to
related, observed variables (e.g., test scores, Likert responses,
etc), which has led to increased popularity of SEMs. SEMs are also related to research on causality and directed acyclic graphs \citep[e.g., ][]{pea13}, to generalized linear mixed models \citep[e.g.,][]{lme4,gelcar13,str13}, and to time series models \citep[e.g.,][]{drioud17}, illustrating the models' broad applicability across disciplines.

A defining feature of the SEM framework is the ability to instantiate regressions on latent variables, as opposed to observed variables. This framework is more general than the traditional mixed modeling framework, allowing for products between latent variables and other free parameters, as might be seen in factor analysis \citep[e.g.,][]{bol89,merwan18}. The generality of SEM implies that the estimation methods are relatively complex, which has historically led researchers to rely on closed-source implementations of optimization methods via software like \proglang{Mplus} \citep{Mplus}, \proglang{LISREL} \citep{jorsor97}, and \proglang{EQS} \citep{EQS}. A small number of more recent \proglang{R} \citep{Rprog} packages, including \pkg{sem} \citep{sempack}, \pkg{OpenMx} \citep{boknea11}, and \pkg{lavaan} \citep{ros12}, provide open source SEM functionality that utilize classical estimation methods including maximum likelihood or least squares.

While maximum likelihood and least squares methods are most popular, Bayesian approaches to SEM and related models have received increased recent attention \citep[e.g.,][]{depvan17,fox10,jac09,kap14,kru11,macedw12,merwan18,mutasp12,vanmul18}. 
Researchers have specifically found the methods to be useful for estimation of complex SEMs \citep[including, e.g., latent variable interactions;][]{leeson07}, for automatically handling uncertainty associated with latent variable estimation, and for scaling to high-dimensional datasets.

Despite the increased popularity of Bayesian latent variable models, coding the models via \proglang{JAGS} \citep{plu03} or \proglang{Stan} \citep{stan} syntax can be difficult, and the resulting sampling can be time-consuming. These issues have been partially addressed by \proglang{R} package \pkg{blavaan} \citep{merros18}, which uses \pkg{lavaan} model specification syntax and originally relied on \proglang{JAGS} for model estimation \citep[via package \pkg{runjags}, which provides an \proglang{R} interface to \proglang{JAGS}; see][]{den15}.
Other \proglang{R} packages have addressed these issues for related models, including \pkg{brms} \citep{bur17} for mixed and multivariate models, \pkg{ctsem} \citep{drioud17} for time series models, \pkg{edstan} \citep{edstan} for item response models, and \pkg{pcFactorStan} \citep{pcfactor} for pairwise comparison factor models.

The original \pkg{blavaan} approach was similar to the \pkg{brms} approach for generalized linear mixed (and related) models, where \proglang{JAGS} code was generated at runtime from the user-specified model syntax. However, this approach became very slow for some models, forcing the user to wait hours or more for enough samples to make inferences. This reduced the viability of \pkg{blavaan} for applied data analysis and for simulation studies, leading us to implement \proglang{Stan} functionality in \pkg{blavaan}. The original \proglang{Stan} implementation was similar to the \proglang{JAGS} implementation, generating \proglang{Stan} syntax for a user-specified model and relying on package \pkg{rstan} \citep{rstan} for MCMC. This \proglang{Stan} implementation has not been formally described, which represents one contribution of the current paper. In general, though, the \proglang{Stan} implementation was not much faster or more efficient than the \proglang{JAGS} approach. 

The primary contribution of this paper is to describe and illustrate a new approach to \proglang{Stan} SEM estimation, which greatly improves the speed and efficiency of model estimation. The approach can be flexibly applied to models in the traditional SEM framework, with general (possibly non-conjugate) prior distributions. It has been implemented in \pkg{blavaan} alongside the previous \proglang{JAGS} and \proglang{Stan} implementations, allowing for easy comparison across implementations. 

In the sections below, we first formally define the models under consideration. We then describe the three MCMC approaches that are now implemented in \pkg{blavaan}: the original \proglang{JAGS} approach described in \cite{merros18}, the original \proglang{Stan} approach that is being formally described in this paper for the first time, and the new \proglang{Stan} approach that is the primary focus of this paper. After describing the approaches, we explicitly discuss some problematic issues associated with estimation of SEMs via MCMC. These are issues that are often overlooked in the literature, but they are necessary to have fully functional Bayesian SEM software. Finally, we compare the three approaches via three examples, highlighting the advantages of the new \proglang{Stan} approach.

\section{Model definition}
The \pkg{blavaan} package generally relies on the \pkg{lavaan} representation of a structural equation model, which is based on the LISREL ``all-y'' representation \citep[e.g.,][]{jorsor97}.

Let $\bm{y}_i$ be the $p$ (continuous) observed variables associated with observation $i$. Then a structural equation model with $m$ latent variables may be represented by the equations
\begin{align}
  \label{eq:sem1}
    \bm{y}_i &= \bm{\nu} + \bm{\Lambda} \bm{\eta}_i +
    \bm{\epsilon}_i \\
  \label{eq:sem2}
    \bm{\eta}_i &= \bm{\alpha} + \bm{B}\bm{\eta}_i + \bm{\zeta}_i,
\end{align}
where $\bm{\eta}_i$ is an $m \times 1$ vector containing the latent 
variables; $\bm{\epsilon}_i$ is a $p \times 1$ vector of residuals; and $\bm{\zeta}_i$ is an $m \times 1$ vector of
residuals associated with the latent variables. The vectors $\bm{\nu}$ and $\bm{\alpha}$ contain intercept parameters for the manifest and latent variables, respectively; $\bm{\Lambda}$ is a
matrix of factor loadings; and
$\bm{B}$ contains parameters
that reflect directed paths between latent variables.

The residuals $\bm{\epsilon}_i$ and $\bm{\zeta}_i$ are assumed to be multivariate normal:
\begin{align}
  \label{eq:mres}
  \bm{\epsilon}_i &\sim N_p(\bm{0}, \bm{\Theta}) \\
  \label{eq:lres}
  \bm{\zeta}_i &\sim N_m(\bm{0}, \bm{\Psi}),
\end{align}
where the associated covariance matrices are often diagonal.
These assumptions imply that the marginal distribution of $\bm{y}$ (integrating out the latent variables) is multivariate normal with parameters
\begin{align}
  \bm{\mu} &= \bm{\nu} + \bm{\Lambda \alpha} \\
  \label{eq:margcov}
\bm{\Sigma} &= \bm{\Lambda}(\bm{I} - \bm{B})^{-1} \bm{\Psi} (\bm{I} - \bm{B}^\top)^{-1} \bm{\Lambda}^\top + \bm{\Theta},
\end{align}
which requires that $(\bm{I} - \bm{B})$ be invertible.
The traditional LISREL framework includes additional matrices for exogenous observed variables, but these are not often utilized in \pkg{lavaan}. Instead, an exogenous observed variable is ``upgraded'' to latent variable status, where the latent variable accounts for all of the observed variable's variance (and the associated variance parameter in $\bm{\Theta}$ is fixed to 0).

Many Bayesian approaches to SEM estimation rely on sampling the $\bm{\eta}_i$ in tandem with other model parameters. This is advantageous because observed variables are often independent conditioned on the $\bm{\eta}_i$, so that the conditional distribution of each observed variable is a univariate normal. However, as we will see later, the sampling of the $\bm{\eta}_i$ can have a major impact on the speed and efficiency of MCMC estimation.

\section{MCMC approaches}
In the sections below, we briefly describe the three MCMC approaches implemented in \pkg{blavaan}: the original \proglang{JAGS} approach, the original \proglang{Stan} approach, and the new \proglang{Stan} approach that is the focus of this paper.

\subsection{Parameter expansion in JAGS}
In previous work \citep{merros18}, we developed a parameter expansion approach that can be applied to SEMs for continuous data \citep[also see][]{paldun07}. The method allows researchers to place prior distributions on intuitive sets of parameters and can be generally implemented in \proglang{JAGS}.

The approach involves converting the model of interest (the ``inferential model'') to an over-parameterized, equivalent model (the ``working model'') from which it is easier to sample. The conversion focuses on a model's covariance parameters, converting each covariance to a ``phantom'' latent variable. This conversion makes observed variables conditionally independent of one another (conditioned on latent variables), meaning that our likelihood involves a series of univariate distributions instead of a single multivariate distribution. Such a conversion can speed up sampling in \proglang{JAGS}, where computations involving the multivariate normal distribution are very slow.
The full details underlying these procedures can be found in \cite{merros18}.

\subsection{Likelihood simplification in Stan}
The phantom latent variable approach is not essential in \proglang{Stan} and, in testing, we found that the approach did not lead to gains in sampling speed or efficiency. However, we did make initial progress in \proglang{Stan} by capitalizing on the structure of the SEM latent variable covariance matrix $\bm{\Psi}$. This capitalization was inspired by related work on estimating multivariate autoregressive models in Stan \citep{jos16}.

We provide an overview of this approach here. For traditional SEMs, the distribution of latent variables can typically be expressed as (see Equation~\eqref{eq:sem2})
\begin{equation*}
\bm{\eta} \sim N((\bm{I} - \bm{B})^{-1} \bm{\alpha}, (\bm{I} - \bm{B})^{-1} \bm{\Psi} (\bm{I} - \bm{B}^\top)^{-1}).
\end{equation*}
Evaluation of this multivariate normal log-likelihood is time-consuming in Stan because we need to compute the inverse and determinant of the covariance matrix. However, for many models, the structure of the covariance matrix leads to simplifications. For example, the matrix $\bm{B}$ is often triangular with zeros along its diagonal (leading to a so-called ``recursive'' model), and the matrix $\bm{\Psi}$ is often diagonal. When both of these properties are fulfilled, we can use standard matrix properties \citep[e.g.,][]{peter08} to write the determinant as a product of scalar values:
\begin{align*}
\text{det}(\bm{I} - \bm{B})^{-1} \bm{\Psi} (\bm{I} - \bm{B}^\top)^{-1}) &= (\text{det}(\bm{I} - \bm{B}))^{-1} \text{det}(\bm{\Psi}) (\text{det}(\bm{I} - \bm{B}))^{-1} \\
 &= 1 \cdot \prod_{i=1}^m \psi_{ii} \cdot 1.
\end{align*}
Relatedly, the inverse of $\bm{\Sigma}$ is simplified as
\begin{equation*}
((\bm{I} - \bm{B})^{-1} \bm{\Psi} (\bm{I} - \bm{B}^\top)^{-1})^{-1} = (\bm{I} - \bm{B}^\top) \bm{\Psi}^{-1} (\bm{I} - \bm{B}),
\end{equation*}
which completely removes the need to compute matrix inversions when $\bm{\Psi}$ is diagonal. When either $\bm{B}$ is triangular or $\bm{\Psi}$ is diagonal (but not both), we can use a subset of the above simplifications to improve sampling efficiency as much as possible. Given a specific model, package \pkg{blavaan} automatically determines which simplifications are available and uses them for \proglang{Stan} estimation. The simplifications are implemented in \proglang{Stan} as a custom log-probability density function. This implementation is available in \pkg{blavaan} via the argument \code{target = "stanclassic"}.

\subsection{New Stan approach}
Both methods mentioned above exploit the fact that the latent variables in the model can be sampled along with other model parameters. This generally simplifies likelihood computations and allows us to immediately extend the methods to situations where observed variables have non-normal distributions. Most Bayesian approaches to SEM, and to other models with ``random'' parameters, sample the latent variables.

However, the sampling of latent variables greatly increases the dimension of the parameter space, which can reduce sampling speed and efficiency. Unlike \proglang{JAGS}, the key to fast sampling in \proglang{Stan} is to work with a model likelihood that is marginal over latent variables. This is somewhat unintuitive, because previous researchers have focused on the simplifications that we can gain from sampling the latent variables. One concern related to using the marginal likelihood involves our inability to make inferences about the latent variables (because they are integrated out of the likelihood). But this concern is addressed by \pkg{blavaan} because, conditional on the other model parameters, the latent variable posterior distribution is tractable. Thus, the latent variables can be sampled in a ``generated quantities'' block within the \proglang{Stan} syntax, even though they do not directly play a role in the MCMC sampling.

The new \pkg{blavaan} approach utilizes the marginal likelihood, and we have written a single \proglang{Stan} program that can estimate the majority of multivariate normal SEMs that a user could specify. This file is compiled once during (or before) package installation. Then, once the user specifies a model, many pieces of information about the data and about the model are passed to the compiled model, with sampling occurring immediately. It is inconvenient to enter all the required model information by hand. Thus, to complement the \proglang{Stan} file, we have new \proglang{R} code that serves as a pipeline from \pkg{blavaan} to the \proglang{Stan} model and back. The \pkg{blavaan} user will not notice many differences, because the commands for model specification and estimation are the same as before. However, the model is now sent to the pre-compiled \proglang{Stan} code by default, whereas the previous approaches wrote \proglang{JAGS} or \proglang{Stan} code at runtime.

It is worth noting that our \proglang{Stan} SEM file stands on its own, so that users of languages beyond \proglang{R} (e.g., \proglang{Python}) could also utilize the file if they can pass all the required data in to the \proglang{Stan} model. This is more challenging than it may sound due to the many pieces of data that are required, including the dimensions of all SEM matrices, the free entries of SEM matrices, equality constraints on free parameters, prior distribution parameters, and so on.

Because our \proglang{Stan} model is precompiled, the possible models that can be estimated are restricted in two ways. First, there is some inflexibility in choice of prior distributions. For most types of model parameters, the form of each parameter's prior cannot be changed (though the specific prior hyperparameters can). For example, regression parameters ($\bm{B}$) in \pkg{blavaan} currently have $\text{N}(0, 10)$ priors by default, where the normal distribution is parameterized by standard deviation. Users can change the mean or standard deviation of this normal prior, but they cannot change the fact that the prior is normal. However, for scale parameters, users have the option to place priors on variances, standard deviations, or precisions. 

The second restriction of the precompiled \proglang{Stan} model involves equality constraints. While our code currently allows for equality constraints within a class of parameters (e.g., loadings can be constrained equal to one another or intercepts can be constrained equal to one another), it does not allow for constraints across classes of parameters. Additionally, if users wish to set one parameter equal to a function of other parameters, that is not currently possible. However, if users desire features that are not included in our current implementation, they can take our \proglang{Stan} file, make the desired changes, and recompile the model. Alternatively, they could use the original MCMC methods available in \pkg{blavaan}, which provide more flexibility because they are not precompiled.

\section{Challenging issues}
While the above methods can be readily applied to ``vanilla'' models such as confirmatory factor analysis with uncorrelated factors, the SEM framework includes many model and data characteristics that require further attention for estimation. Below, we highlight three characteristics requiring special attention.

\subsection{Covariance parameters}
The general SEM presented earlier includes two covariance matrices with free parameters: $\bm{\Theta}$ and $\bm{\Psi}$. These matrices can include some fixed values and some free values, so prior distributions for these matrices are not straightforward. That is, there are some models for which we cannot simply place an inverse Wishart prior on the covariance matrix, nor an LKJ prior \citep{lkj09}. Those priors were meant for unrestricted covariance/correlation matrices, not for matrices with some fixed values and some free values. 

In the \proglang{Stan} approaches, we consequently decompose the covariance matrices into standard deviations and correlations. For example, $\bm{\Theta}$ is written as
\begin{equation}
  \bm{\Theta} = \bm{D}_\Theta \bm{R}_\Theta \bm{D}_\Theta,
\end{equation}
where $\bm{D}_\Theta$ is a diagonal matrix of standard deviations and $\bm{R}_\Theta$ is a correlation matrix. Prior distributions are then placed individually on the free standard deviation parameters and on the free correlation parameters within the two matrices. This approach is similar to that of \cite{barmcc00}, and \cite{liuzha16} provide a comparison of this approach to the use of inverse Wisharts in the context of growth curve models. 

The use of an independent prior on each free parameter can sometimes lead to a non-positive definite covariance matrix during MCMC sampling. \proglang{Stan} is able to reject such a covariance matrix and continue sampling, whereas \proglang{JAGS} will terminate. This is why we developed the parameter expansion method in \proglang{JAGS}: the parameter-expanded model involves diagonal covariance matrices that cannot become non-positive definite. But the non-positive definite covariance matrices still have implications for model calibration, as detailed in the simulation-based calibration study later.

We are aware of a variety of other prior distributions proposed for covariance matrices \cite[e.g.,][]{chugel15,con03,mulper18,spe18}. The strategy implemented in \pkg{blavaan} is worthwihle because it is relatively easy to specify informative prior distributions for individual standard deviation and correlation parameters in the model. In contrast, many of the other prior distributions are proposed for convenience or due to the fact that they maintain positive definiteness, and they have less-intuitive interpretations as compared to our approach. We plan to further consider these alternative priors in the future.

\subsection{Missing data}
While it is often useful and desirable to directly model the missing values with the rest of the model \citep[e.g.,][]{mer11,omumou99}, \pkg{blavaan} employs a ``missing at random'' approach to missing data that differs across \proglang{JAGS} and \proglang{Stan}. In \proglang{JAGS}, one can include NA values in the data, and \proglang{JAGS} will sample these missing values as if they were extra model parameters. In contrast, \proglang{Stan} does not allow NA values in the data, so that one must handle the missing data manually.  We utilize a ``full information'' likelihood \citep[e.g.,][]{wot00} in our \proglang{Stan} models, which is the same likelihood that is used to handle missing data in \pkg{lavaan} and other software that performs maximum likelihood SEM estimation. This requires some additional overhead in preparing the data to be sent to \proglang{Stan}, because each case's observed values must be indexed, and cases are sorted by missing data pattern to speed up computations. Missing values could also be directly sampled (``imputed'') in \proglang{Stan}, though this functionality is not currently available.

\subsection{Latent variable scaling}
Structural equation models typically require some parameter constraints to achieve parameter identification, where we must ``set the scale'' of each latent variable. The two most popular ways to do this involve (i) fixing each latent variable's variance to 1, or (ii) fixing a single loading (parameter in $\bm{\lambda}$) to 1 for each latent variable. Of these two, the latter method is most straightforward to implement in a Bayesian setting.

The former method (of fixing each latent variance to 1) is more challenging. This is because, as described by \cite{pee12}, one loading per latent variable must be sign constrained to achieve global parameter identification. Otherwise, the sign of each loading may flip back and forth, with a model's regression parameters and covariance parameters potentially flipping along with the loadings. One solution to this issue involves the placement of a truncated normal prior (truncated from below at 0) on one loading per latent variable, preventing the sign changes. This solution is adopted in \pkg{blavaan}'s \proglang{JAGS} approach to model estimation.

A different solution is implemented in \pkg{blavaan}'s \proglang{Stan} approaches. In those approaches, the sign flipping is allowed to occur during MCMC sampling. The issue is then handled after sampling, in the ``generated quantities'' block. In this block, one loading per latent variable is transformed to always be positive, and the signs of associated parameters (loadings, regressions, and covariance parameters) are flipped every time the sampled value of the focal loading is negative. This approach can improve sampling efficiency because no boundary constraints are introduced in the parameter space. This approach was discussed in a thread on the Stan Discourse site (\url{https://discourse.mc-stan.org/t/latent-factor-loadings/1483}).

\section{Applications}
The estimation approaches described above are all implemented in package \pkg{blavaan} for general SEM estimation. These include the original \proglang{JAGS} approach (obtained via argument \code{target = "jags"}), the original \proglang{Stan} approach (\code{target = "stanclassic"}), and the new \proglang{Stan} approach (\code{target = "stan"}).

For example, the following code specifies a model for the well-known ``political democracy'' data \citep{bol89} and estimates it via each of the three approaches. 
This dataset includes 75 countries measured on 11 attributes, seven of which were measured in 1960 and 4 of which were measured in 1965. The intent of the model is to study relationships between countries' levels of industrialization and democracy over time.

\begin{Schunk}
\begin{Sinput}
R> model <- '
+    # measurement model
+      ind60 =~ x1 + x2 + x3
+      dem60 =~ y1 + y2 + y3 + y4
+      dem65 =~ y5 + y6 + y7 + y8
+    # regressions
+      dem60 ~ ind60
+      dem65 ~ ind60 + dem60
+    # residual correlations
+      y1 ~~ y5
+      y2 ~~ y4 + y6
+      y3 ~~ y7
+      y4 ~~ y8
+      y6 ~~ y8
+  '
R> fit1 <- bsem(model, data = PoliticalDemocracy, target = "jags")
R> fit2 <- bsem(model, data = PoliticalDemocracy, target = "stanclassic")
R> fit3 <- bcfa(model, data = PoliticalDemocracy, target = "stan")
\end{Sinput}
\end{Schunk}

The above commands use the default number of burnin/warmup and sampling iterations, as well as the package's default prior distribution for each type of model parameter. The default prior distributions have generally been chosen to be weakly informative for a variety of SEMs typically encountered in practice; some further discussion of prior distributions appears later in the simulation-based calibration section.

Following model estimation, convergence diagnostics such as Rhat and effective sample size are immediately available via the \code{summary} method and \code{blavInspect()} function, and many types of plots are available via the \code{plot} method, which relies on package \pkg{bayesplot} \citep{bayesplot}. Further examples of \pkg{blavaan} syntax and functionality can be found in \cite{merros18}, noting that \code{target="jags"} was the default at the time that paper was written, while \code{target="stan"} is now the default.

In the following sections, we compare the three MCMC approaches on speed and sampling efficiency via three example models. We then study the extent to which the best method (the marginal \proglang{Stan} method) is calibrated.

\subsection{Performance Comparison}
All comparisons are carried out on a Dell desktop with a large amount of RAM, running Ubuntu Linux. 
We define sampling efficiency as ``effective sample size per second'' (ESS/s). Effective sample sizes are computed via the \pkg{rstan} \code{monitor()} function, and sampling time is measured after \proglang{Stan} model compilation. The warmup time for \proglang{Stan} models was fixed to 300 iterations, whereas the burn-in time for \proglang{JAGS} models was fixed to 1000 iterations. There is arbitrariness in the warmup and burn-in choices, so that the ESS/s metric is somewhat crude. But we think the metric is sufficient to illustrate the advantage of the new \proglang{Stan} approach.

We examine the MCMC methods' speed and efficiency on three models, two of which are popular models often used to illustrate SEM methods. The third is a more complex model that is known to pose difficulties for the original \pkg{blavaan} approach. We do {\em not} conduct a full Monte Carlo study here, so our results are subject to noise. But the results are generally consistent across the models presented here as well as many others not presented, so we think they can be taken as general evidence for the approaches' relative performance. The results are also consistent with those of \cite{yacdod20}, who study marginalization of discrete latent variables in ecological models.


\subsubsection{Political democracy}
\begin{figure}
\begin{Schunk}

{\centering \includegraphics[width=4in]{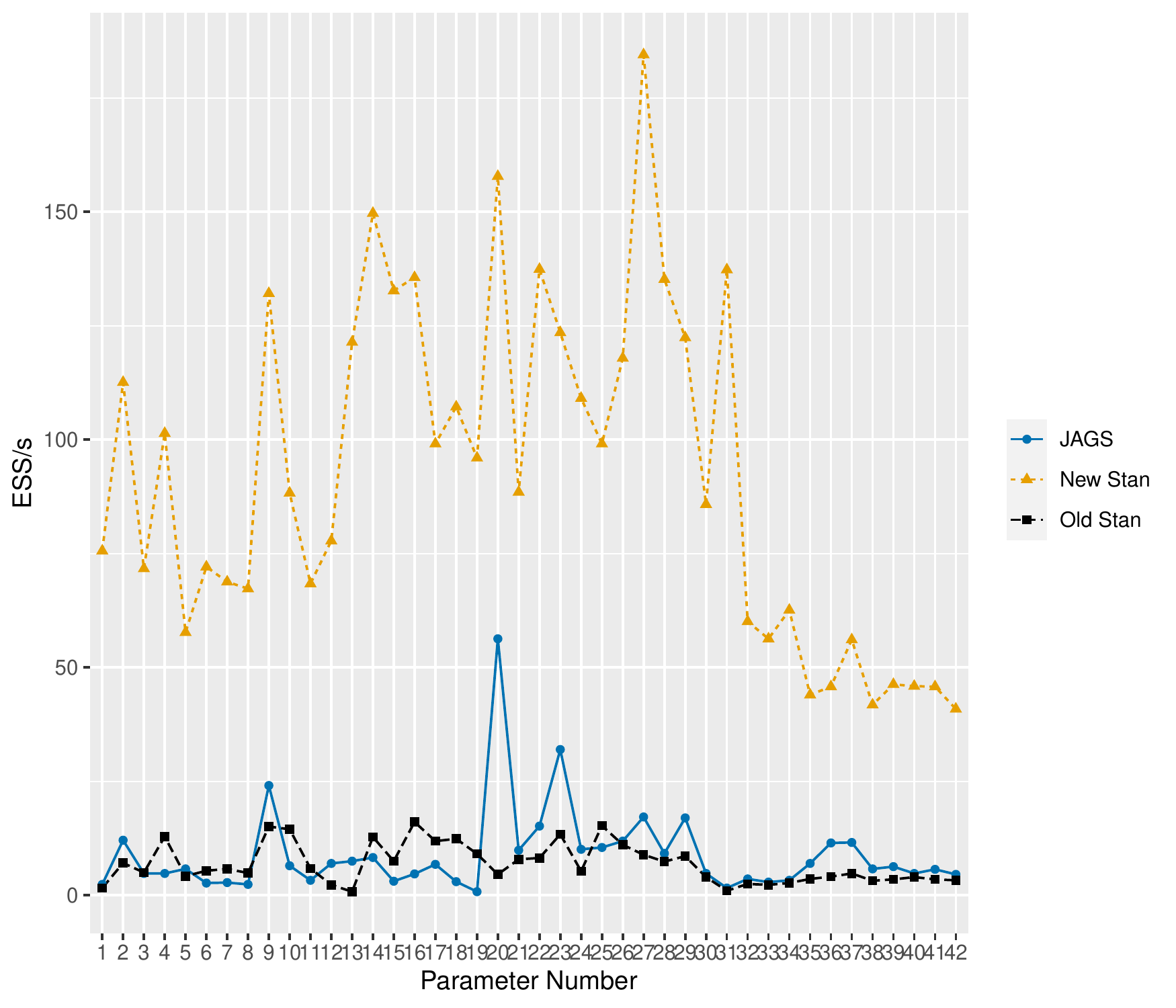} 

}

\end{Schunk}
  \caption{Sampling efficiency of the MCMC procedures for the Bollen example. Parameter numbers correspond to different types of parameters: loadings are 1--8; regressions are 9--11; observed variances are 12--22; latent variances are 23--25; intercepts are 26--36.}
  \label{fig:pd}
\end{figure}

For our first comparison, we continue with the \cite{bol89} political democracy model. The \pkg{blavaan} code to fit the model was shown earlier.
The \proglang{JAGS} method was fastest here, averaging 0.55 seconds per 100 iterations. Next fastest was the new (marginal) \proglang{Stan} method, averaging 1.44 seconds per 100 iterations, followed by the old \proglang{Stan} method at 
7.36 seconds per 100 iterations. But it is more important to examine the methods' sampling efficiencies (effective sample size per second), which are shown in Figure~\ref{fig:pd}. A separate metric is shown for each parameter, with the parameters being numbered along the x-axis. Parameters are ordered on the x-axis by parameter type, with the details being shown in the figure caption. The figure shows that the speed of \proglang{JAGS} is offset by the effective sample size, so that the new \proglang{Stan} method is best in terms of sampling efficiency.
The old \proglang{Stan} method exhibits efficiency similar to that of \proglang{JAGS}.

\subsubsection{Holzinger and Swineford}
Our second example involves a confirmatory factor analysis of the \cite{holswi39} data. This is the version of the dataset included in package \pkg{lavaan}, which has 301 individuals measured on nine cognitive scales. The confirmatory factor model fit to the data includes three latent variables, each of which is associated with three observed variables. The \pkg{blavaan} code to specify and fit the model is
\begin{Schunk}
\begin{Sinput}
R> HS.model <- ' visual  =~ x1 + x2 + x3
+                textual =~ x4 + x5 + x6
+                speed   =~ x7 + x8 + x9 '
R> fit <- bcfa(HS.model, data = HolzingerSwineford1939)
\end{Sinput}
\end{Schunk}
where additional arguments would typically be used to specify the number of sampling iterations, to specify priors, to specify the MCMC sampler, and so on. 

\begin{figure}
\begin{Schunk}

{\centering \includegraphics[width=4in]{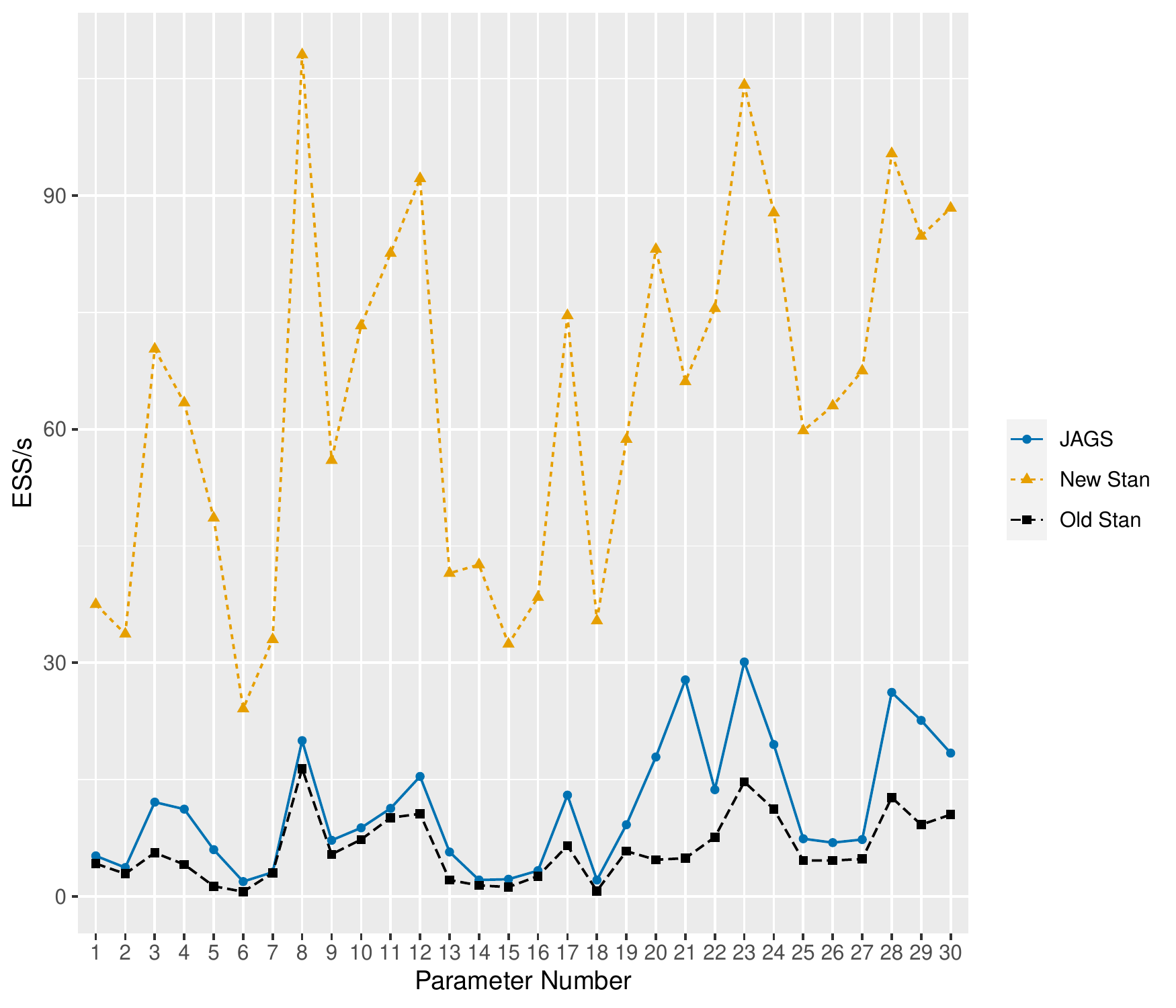} 

}

\end{Schunk}
  \caption{Sampling efficiency of the MCMC procedures for the Holzinger \& Swineford example. Parameter numbers correspond to different types of parameters: loadings are 1--6; observed variable variances are 7--15; latent variable variances are 16--18; latent variable covariances are 19--21; intercepts are 22--30.}
  \label{fig:hs}
\end{figure}

In terms of speed, the \proglang{JAGS} method is again fastest, averaging 0.91 seconds per 100 iterations. This was followed by the marginal \proglang{Stan} method at 2.83 seconds per 100 iterations, then the old \proglang{Stan} method at 9.55 seconds per 100 iterations.
The ESS/s metrics for this model are visualized in Figure~\ref{fig:hs}. The graph is similar to that of the previous section, with fewer parameters in this model as compared to the last model. We see that the gold line, representing the new \proglang{Stan} method, is the highest for all the model parameters, being at least twice as large as the other methods' efficiencies. The \proglang{JAGS} and old \proglang{Stan} methods are again similar to one another for this example, with \proglang{JAGS} being better for the majority of parameters.

\subsubsection{Growth model}
For our final comparison, we use a ``Multiple indicator univariate latent change score'' model presented in \cite{kiebra18}. The \pkg{blavaan} code to fit this model, as specified by \cite{kiebra18}, is
\begin{Schunk}
\begin{Sinput}
R> MILCS <- '
+    COG_T1 =~ 1*T1X1 + T1X2 + T1X3
+    COG_T2 =~ 1*T2X1 + equal("COG_T1 =~ T1X2")*T2X2 + equal("COG_T1 =~ T1X3")*T2X3
+  
+    COG_T2 ~ 1*COG_T1
+    dCOG1 =~ 1*COG_T2
+    COG_T2 ~ 0*1
+    COG_T2 ~~ 0*COG_T2
+  
+    dCOG1 ~ 1
+    COG_T1 ~ 1
+    dCOG1 ~~ dCOG1
+    COG_T1 ~~ COG_T1
+    dCOG1 ~ COG_T1
+  
+    T1X1 ~~ T2X1
+    T1X2 ~~ T2X2
+    T1X3 ~~ T2X3
+  
+    T1X1 ~~ T1X1
+    T1X2 ~~ T1X2
+    T1X3 ~~ T1X3
+  
+    T2X1 ~~ equal("T1X1 ~~ T1X1")*T2X1
+    T2X2 ~~ equal("T1X2 ~~ T1X2")*T2X2
+    T2X3 ~~ equal("T1X3 ~~ T1X3")*T2X3
+  
+    T1X1 ~ 0*1
+    T1X2 ~ 1
+    T1X3 ~ 1
+    T2X1 ~ 0*1
+    T2X2 ~ equal("T1X2 ~ 1")*1
+    T2X3 ~ equal("T1X3 ~ 1")*1
+  '
R> fit <- blavaan(MILCS, data = simdatMILCS, fixed.x = FALSE)
\end{Sinput}
\end{Schunk}
where further information on this model and its specification can be found in the original authors' paper. 
We fit the model to 500 artificial observations, where the data were generated via code that was included with the Kievit paper. 
This model, and others described in the Kievit paper, have been especially difficult to fit in \pkg{blavaan}, requiring long run times and high autocorrelation among parameter draws. We ended up thinning the \proglang{JAGS} samples by 20 in our analyses here, because it was the only way that we could consistently obtain an Rhat value below 1.05.

The sampling speed is now reversed, with the marginal \proglang{Stan} method at 23.4 seconds per 100 iterations, the \proglang{JAGS} method at 28.4 seconds per 100 iterations, and the old \proglang{Stan} method at 577.27 seconds per 100 iterations. If we instead compute the \proglang{JAGS} speed while accounting for thinning (i.e., counting only each twentieth iteration in the computations), then the \proglang{JAGS} speed is at 567.98 seconds per 100 iterations.

\begin{figure}
\begin{Schunk}

{\centering \includegraphics[width=4in]{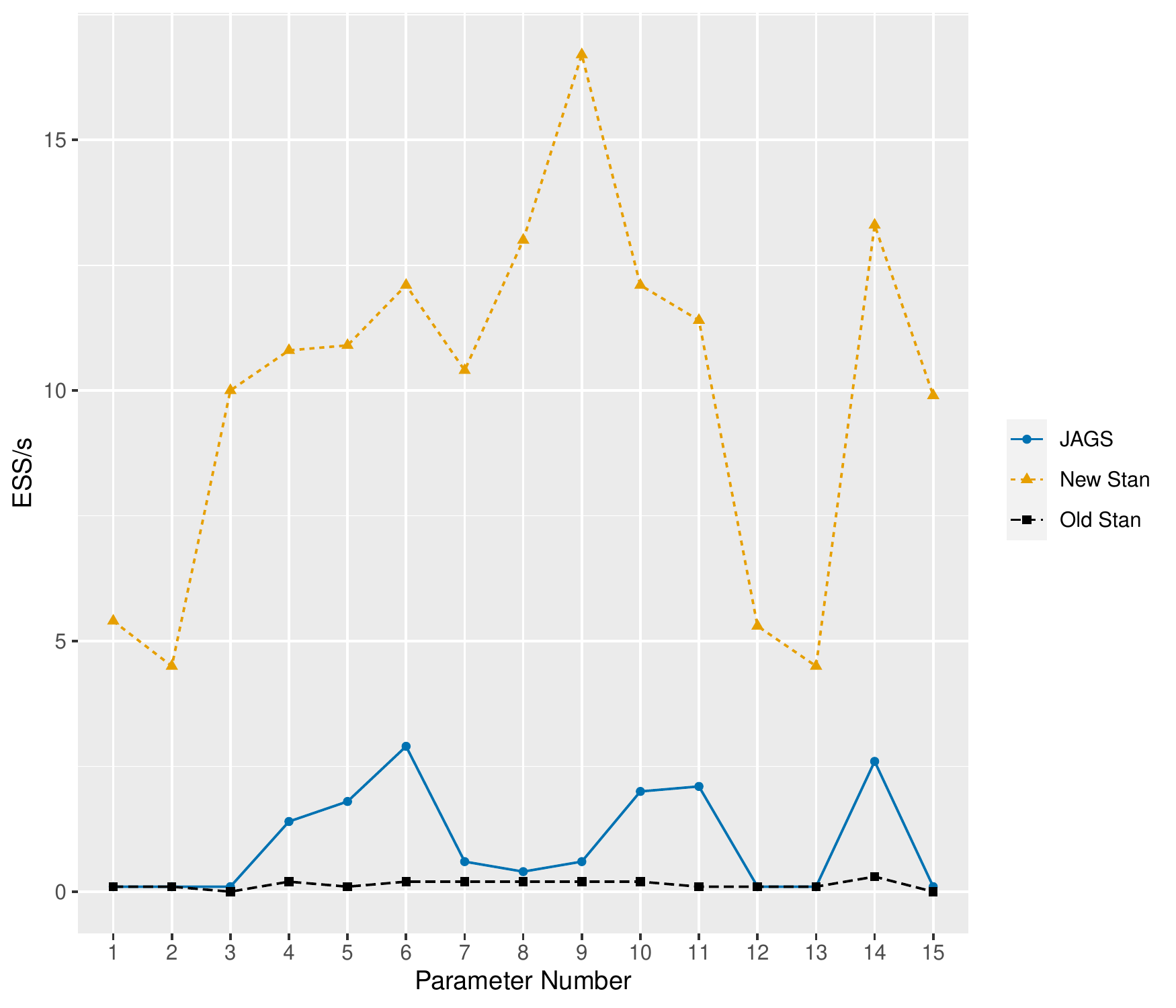} 

}

\end{Schunk}
  \caption{Sampling efficiency of the MCMC procedures for the growth model example. Parameter numbers correspond to different types of parameters: loadings are 1--2; regressions are 3; observed variances are 4--6; observed variable covariances are 7--9; latent variances are 10--11; intercepts are 12--13; latent means are 14--15.}
  \label{fig:gr}
\end{figure}

The approaches' sampling efficiencies are shown in Figure~\ref{fig:gr}, where parameter ordering is again described in the figure caption. The \proglang{JAGS} and old \proglang{Stan} methods are very low for this model, with the new \proglang{Stan} method displaying much better efficiency and yielding useful results in a matter of minutes, as opposed to hours or days. This and related examples (not shown) convinced us to make the new \proglang{Stan} method the default in \pkg{blavaan}, replacing the original default of \proglang{JAGS}. The new \proglang{Stan} method reliably produces fast, efficient samples for a large number of models, whereas the other methods exhibit more variability in their speeds and efficiencies, and are seldom clearly better than the new \proglang{Stan} method.

\subsection{Verification and Simulation-based Calibration}
The posterior estimates resulting from \pkg{blavaan} have been verified in a few manners. Initially, we treated \pkg{lavaan} as a gold standard, comparing posterior means and standard deviations under weak priors to the maximum likelihood estimates and standard errors from \pkg{lavaan}. For all three MCMC methods, the posterior means obtained under this approach typically agree with \pkg{lavaan} estimates to about one decimal point. The posterior standard deviations tend to be close to, but slightly larger than, the maximum likelihood standard errors. Now that there are multiple MCMC methods implemented in \pkg{blavaan}, we have also been able to compare MCMC methods to one another in order to verify that they were producing similar posterior distributions.

Here, we use the simulation-based calibration method proposed by \cite{talts2018} to study the calibration of \pkg{blavaan}'s new (marginal) \proglang{Stan} implementation. This involved repeatedly generating data from the model's prior distribution, fitting the model to the generated data, and examining the ranks of the posterior MCMC samples relative to samples from the prior distribution. If the MCMC algorithm is calibrated, then these ranks should be approximately uniformly distributed. Deviations from uniformity are then taken as miscalibration.

\begin{table}
  \footnotesize
  \label{tab:sbcpri}
  \begin{center}
  \begin{tabular}{lcccccc} \hline
                    & $\nu$ & $\lambda$ & $\beta$ & $\theta$ & $\psi$ & $\rho$ \\\hline
    Set 1 (default) & N(0,32) & N(0,10) & N(0,10) & Gamma(1,.5) & Gamma(1,.5) & Beta(1,1) \\
    Set 2 (informative) & N(0,32) & N(1.25,.25) & N(1.5,.25) & Gamma(10,10) & Gamma(10,10) & Beta(5,5) \\\hline
  \end{tabular}
  \end{center}
  \caption{Prior distributions used in the simulation-based calibration study. Normal priors are parameterized using standard deviations. Gamma priors are placed on standard deviations associated with $\theta$ and $\psi$ parameters, as opposed to variances or precisions.}
\end{table}

\subsubsection{Method}
Our simulation-based calibration study utilized the political democracy model presented earlier. We generated 500 datasets of size 75 from the prior predictive distribution and fit the model to each generated dataset via MCMC. The study involved two conditions that differed by the prior distributions that were used. First, we used the default prior distributions from \pkg{blavaan}, which are intended to be weakly informative in many situations encountered in practice. Second, we used a set of more informative prior distributions to contrast with the noninformative priors. Both sets of prior distributions are shown in Table~\ref{tab:sbcpri}.

\subsubsection{Results}
\begin{figure}
\begin{Schunk}

{\centering \includegraphics[width=6in]{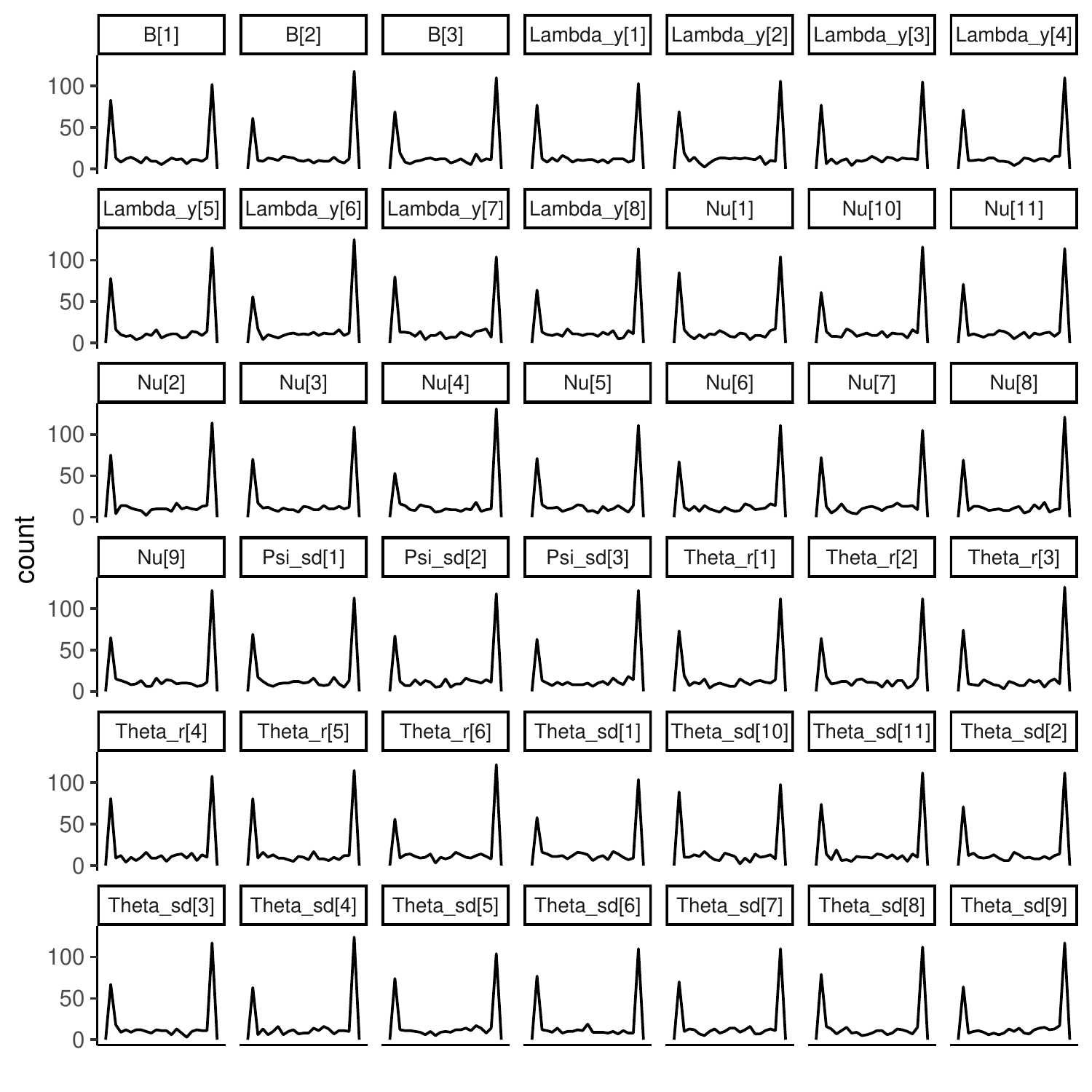} 

}

\end{Schunk}
\caption{Simulation-based calibration rank frequencies, default \pkg{blavaan} priors.}
\label{fig:sbcni}
\end{figure}

Rank frequencies for the \pkg{blavaan} default priors are shown in Figure~\ref{fig:sbcni}. Perhaps surprisingly, the distributions are far from uniform, with peaks generally occurring near zero and one. These peaks represent posterior distributions that exhibit less variability than they should, given the non-informative priors from which we started. Clearly, the results are far from the uniformity that would be expected from a calibrated algorithm.

The non-uniformity occurs because our model has a large number of parameters with independent prior distributions, with the parameters all contributing to the model-implied covariance matrix (see Equation~\eqref{eq:margcov}). For the model considered here, there are many combinations of parameters that lead to a non-positive definite covariance matrix, and the MCMC sampler will avoid these combinations of parameters during sampling. In the simulation-based calibration study, this leads to posterior samples that are not calibrated with respect to the independent priors. Instead, we might say that the posteriors are calibrated with respect to regions of the prior distribution that are positive definite.

\begin{figure}
\begin{Schunk}

{\centering \includegraphics[width=6in]{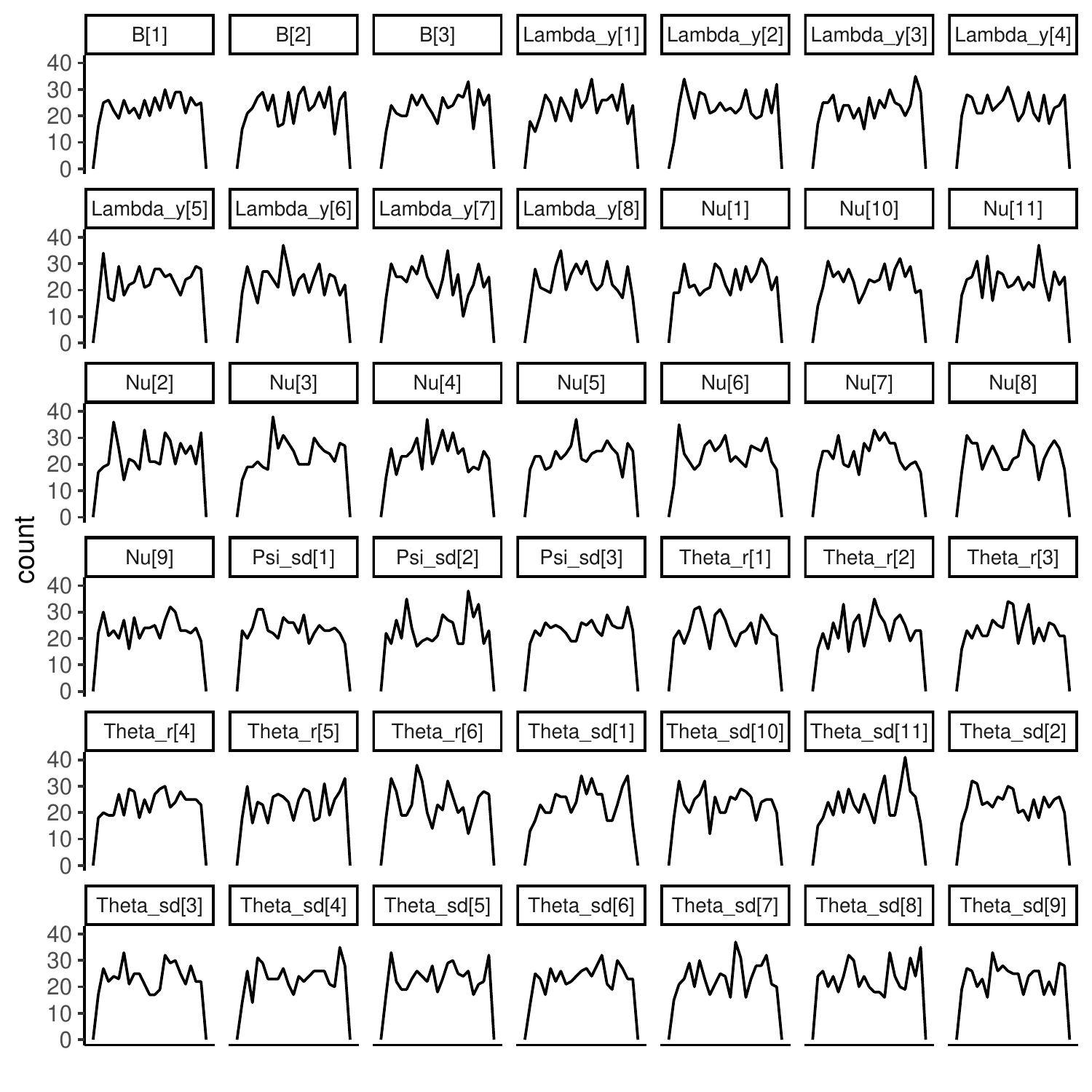} 

}

\end{Schunk}
\caption{Simulation-based calibration rank frequencies, informative priors.}
\label{fig:sbcin}
\end{figure}

To provide evidence that the MCMC algorithm is indeed calibrated with respect to priors that maintain positive definiteness, we show the results of the informative priors in Figure~\ref{fig:sbcin}. These frequencies are now much closer to uniform, because the information in the prior distributions now generally leads to positive definite model covariance matrices. This interplay between the informativeness of prior distributions and posterior calibration is worthy of further attention, because existing MCMC algorithms for SEM use a series of independent priors on parameters that each play a role in the model-implied covariance matrix. A researcher's priors may be more informative than expected, based solely on the fact that the model-implied covariance matrix must remain positive definite during MCMC sampling. Further, depending on the model likelihood used (marginal vs.\ conditional), the degree of information present in uninformative priors may vary. These results highlight the utility of \pkg{blavaan} for conducting detailed study of MCMC algorithms, as well as the fact that there is room to improve the default \pkg{blavaan} priors in the future.

\section{Conclusion}
The results in this paper show that we can improve sampling efficiency by integrating the latent variables out of the model likelihood, which is the opposite of most popular approaches to Bayesian SEM estimation \cite[where the popular approaches are largely based on results summarized by, e.g.,][]{lee07,sonlee12}. We can expect the marginal sampling efficiency to be even more advantageous as sample sizes increase, because the sample size has no impact on the dimension of the parameter space here. In contrast, the dimension of the parameter space increases with sample size under conditional approaches, where latent variables count as parameters.

While the marginal approach is promising, use of the marginal likelihood leads us back to problems that frequentists often encounter in SEM. These problems include the fact that the marginal likelihood does not have a closed form when we have non-normal observed variables (e.g., ordinal variables) or when we have latent variable interactions. We think that some progress can be made here by employing other Bayesian methods, including data augmentation \citep[e.g.,][]{chigre98} in the ordinal case. The use of data augmentation for psychometric models has been described by \cite{fox10} and \cite{foxmul17}, and such methods may be implemented in future versions of \pkg{blavaan}.

For situations where the marginal likelihood does not exist in closed form, it is also possible to move back to the original \pkg{blavaan} approaches that sample the latent variables. However, in our experience, the original approaches are even slower and less efficient in those situations (as compared to the models considered here), making them questionable for applied work. Further, even if those methods did exhibit reasonable efficiency, the marginal likelihood is generally necessary for obtaining suitable information criteria such as DIC \citep{spibes02} or WAIC \citep{wat10}. \cite{merfur19} discuss why the marginal likelihood is preferable here, and \cite{zhang2019} discuss related applications of DIC to multilevel item response models. Thus, we think that use of the new \proglang{Stan} approach, paired with new tricks for handling non-normal observed variables, is the most viable approach for applications to the non-normal modeling situations typically encountered in practice.

We should note that the three methods here are not the only ones that can be conceptualized in \proglang{JAGS} or in \proglang{Stan}. For example, the pre-compiled marginal \proglang{Stan} approach can be modified so that the latent variables are part of the model likelihood. This leads to a method that is similar to the ``old \proglang{Stan}'' method, except that it simplifies computation of the model likelihood in different ways. In limited testing, we found that this approach was somewhat closer to the \proglang{JAGS} approach in sampling efficiency, but still similar to the ``old \proglang{Stan}'' approach studied in this paper. Alternatively, it is possible to define a marginal approach in \proglang{JAGS}, but our limited testing there indicates that use of the multivariate normal distribution leads to major decreases in \proglang{JAGS} sampling efficiency.

Finally, our analyses here have included no use of parallelization. While between-chain parallelization is immediately available in \pkg{blavaan}, there have been recent advances in within-chain parallelization in \proglang{Stan}. These advances are promising for further improvement of \proglang{Stan} sampling efficiency, and we plan to include this in future versions of \pkg{blavaan}.

\section{Acknowledgments}
This work was partially supported by a research leave to the first author, which was provided by the University of Missouri.

\bibliography{refs}
\end{document}